\documentclass[11pt]{article}

\usepackage{graphicx,epsfig,amssymb,amsmath,dsfont}
\begin{document}
\title{Entropy bound and local quantum field theory}
\author{Andreas Aste\\
Department of Physics and Astronomy, Universit\"at Basel\\
Klingelbergstrasse 82, 4056 Basel, Switzerland}
\date{May 29, 2007}
\maketitle
\begin{abstract}
I comment on Ulvi Yurtsever's result,
which states that the entropy of a truncated bosonic Fock space
is given by a holographic bound when the energy of the Fock states
is constrained gravitationally. The derivation given in Yurtsever's
paper contains an subtle mistake, which invalidates the result.
A more restrictive, non-holographic entropy bound is derived.
\vskip 0.3 cm
\end{abstract}
\section{Introduction}
\noindent In \cite{Hooft},'t Hooft presented a simple derivation of
the entropy of a closed spacelike surface, e.g.,
the surface of a sphere with total surface area $A$
and volume $V$, which contains (bosonic) quantum fields.
Considering all possible field configurations inside this
surface, one may ask how many mutually orthogonal quantum field states
can be excited inside the space under consideration.
If the states are to be observable for the outside world, their energy
has to be less than $1/4$ times the diameter of the sphere in natural units,
otherwise the surface would lie within the Schwarzschild radius. The most
probable state would be a gas at some temperature $T=1/\beta$ with an energy
approximately given by
\begin{equation}
E=C_1Z V T^4\, \label{energy}
\end{equation}
where $Z$ is the number of different fundamental particle types (with mass less
than $T$) and $C_1$ a numerical constant of order one, again all in natural units.
The total entropy is $S=C_2Z V T^3$,
where $C_2$ is another dimensionless constant.
Since the Schwarzschild limit requires that
$2E<(V/ \frac{4}{3} \pi)^{\frac{1}{3}}$, one obtains with eq. (\ref{energy})
$T<C_3Z^{-{\frac{1}{4}}}V^{-{\frac{1}{6}}}$,
so that $S<C_4Z^{\frac{1}{4}}V^{\frac{1}{2}}=C_5Z^{\frac{1}{4}}A^{\frac{3}{4}}$.
The $C_i$ are all constants of order 1 in natural units.  Since in
quantum field theories, at sufficiently low temperatures, $Z$ is
limited by a dimensionless number one finds that this entropy is small
compared to the entropy of a black hole which is proportional to the
black hole surface area, if the area $A$ is sufficiently
large (slightly beyond the Planck scale if $Z$ is not extremely
large, to be more precise).

According to Yurtsever's original idea,
imposing an upper bound on the total energy of the corresponding
Fock states which ensures that the system is in a
stable configuration against gravitational collapse
and imposing a cutoff on the maximum energy of the field modes
of the order of the Planck energy
leads to an entropy bound of holographic type.
This idea has been generalized {\it{bona fide}} in \cite{Aste}
to the case of spacetimes with arbitrary dimension.

However, for 't Hooft's classical result that $S \propto A^{\frac{3}{4}}$
to disagree with the conjectured entropy of the truncated Fock space
$\propto \! A$ would require a disagreement between
the microcanonical and canonical ensembles for a system with a large
number of degrees of freedom. Furthermore, the additional
restriction of the allowed states by a cutoff on the maximum
energy of the field modes should rather lead to an even more
restrictive bound on the entropy.

In order to understand that there is an error in the arguments given by Yurtsever,
it is advantageous to focus on eq. (28) in \cite{Yurtsever} first.
There, the (approximate) dimension of the truncated Fock space is
expressed by the function
\begin{equation}
q(z)=\sum \limits_{n=0}^{\infty} \frac{z^n}{(n!)^2}. \label{sum}
\end{equation}
In the sum above, each individual term $d_n=z^n/(n!)^2$ denotes
basically the number of orthogonal quantum field states
of a massless bosonic field with the property
that exactly $n$ arbitrary but different modes are excited
(see also eq. (21) and (27) in \cite{Yurtsever}).
This number is related in \cite{Yurtsever} to a number of points lying on subpolyhedra.
It is straightforward to see that the dominant contribution to $q(z)$
in eq. (\ref{sum}) stems from terms with $n$ located narrowly around
$n \simeq \sqrt{z}$.

That this is indeed incorrect is most easily illustrated for
the case of a cube $\mathcal{C}$ with side length $L$.
In this case, one has according to \cite{Yurtsever} $z=(\sqrt{3}/4) L^4$
(in units of the Planck length)
if the maximum energy of allowed field modes is given by the Planck energy.
According to the hoop conjecture that
a nonspherically compressed object will form a black hole around
itself when its circumference in all directions becomes less than
the critical circumference \cite{Thorne}, one gets a bound for the
energy of the cube $E < (\sqrt{3}/4)L$. The energy of the ground mode
described by a wave function $\Phi(x,y,z) \sim \sin(\pi x/L) \sin(\pi y/L)
\sin (\pi z/L)$ fulfilling the wave equation $\Box \Phi=0$ and vanishing on
$\partial \mathcal{C}$, is given for massless particles by
$\Omega_1 =\frac{\sqrt{3} \pi}{L}$ and would be even larger for massive particles
(note that the wave functions $\sim \sin(\vec{k} \vec{x}) e^{-i \omega_{\vec{k}} t}$
given by Yurtsever do not vanish on $\partial \mathcal{C}$).
Therefore, the number $d$ of different modes that can be
excited simultaneously must fulfill at least the bound
\begin{equation}
d \Omega_1  < \frac{\sqrt{3} L}{4} \quad \mbox{or} \quad
d < N_{max}=\frac{L^2}{4 \pi}. \label{bound}
\end{equation}
If $d$ is larger than $N_{max}$, then the total energy of the
cube clearly exceeds the gravitational energy bound even if only
lowest energy modes are excited, hence the corresponding
Fock states do not contribute to $q(z)$.
A slightly more detailed analysis presented below leads to the much
more restrictive bound $d < C L^{\frac{3}{2}}$, where $C$ is a numerical constant
of order one, if it is taken into account that except for the ground mode
all other excited modes have an energy which is larger than the energy quantum
$\Omega_1$, but the bound above is already sufficient to highlight the problem.
Since according to Yurtsever's combinatorics,
the dominant contribution to the number of dimensions
of the truncated Fock space comes from terms with $n \simeq \sqrt{z}
=\frac{3^{1/4}}{2} L^2 > N_{max}$, one has a clear contradiction to
the bound eq. (\ref{bound}).

The combinatorial quantity $S_n$ defined by eq. (20) in \cite{Yurtsever} is
evaluated incorrectly and must vanish for $n > C L^{\frac{3}{2}}$.
Although the very simple reasoning presented above is already sufficient to rule out
the entropy bound given in \cite{Yurtsever}, we give a more detailed derivation of a
much stronger bound for the entropy of a scalar quantum field in the
following section. This derivation and notational details follow
closely the original work of Yurtsever, and we will use natural units or
include natural constants $\hbar$ and $c$ where this is also done in \cite{Yurtsever}.

\section{Non-holographic entropy bound}
Before tackling the actual calculations, we point out that the entropy
we intend to calculate is unambiguously defined as a combinatorial quantity
(eq. (15) in \cite{Yurtsever}) which is the dimension of the constrained Fock space
$\mbox{dim} \, \mathcal{H}_F = W(B) \equiv \mbox{number of } (n_1,n_2,...,n_N), \, n_i \in
\mathds{N} $ such that
\begin{equation}
\sum \limits_{i} n_i \Omega_i < B \quad (\mbox{or} \, \,
\sum \limits_{i} n_i \hbar \omega_i < E_{max}),
\label{start}
\end{equation}
where
\begin{equation}
B=\eta \frac{\sqrt{3}}{4} \frac{L}{l_p} \quad
(E_{max}=\frac{\hbar}{\tau_p} B, \, \Omega_i=\tau_p \omega_i),
\end{equation}
according to the hoop conjecture, is the black hole energy bound expressed in units
of the Planck energy $\epsilon_p=\hbar/\tau_p=\hbar c/l_p$ and the $n_i$'s are the
occupation numbers of the corresponding modes containing quanta with energy $\Omega_i$.
Therefore, ambiguities as they may appear in the definition of the thermodynamic entropy do
not appear in the following calculation, which follows the notation of Yurtsever
but corrects an erroneous assumption that was made in \cite{Yurtsever}.

Yurtsever constructs the dimension $W(B)$ of the constrained Fock space
according to
\begin{equation}
W(B)=1+N+\sum \limits_{n=1}^{N} \frac{S_n}{n!}, \quad
N=\frac{L^3}{2 \pi^2 c^3} \int \limits_{0}^{2 \pi \mu / \tau_p} \omega^2 d \omega
=\frac{4 \pi \mu^3}{3} \Biggl( \frac{L}{l_p} \Biggr)^3,
\label{toobigsum}
\end{equation}
where $N$ is the total number of modes with a quantal energy smaller than $2 \pi \mu \hbar/\tau_p$
(eq. (3) in \cite{Yurtsever}). 
We point out again what the meaning of the terms $S_n/n!$ actually is. $S_1$ is the dimension of
the subspace where exactly one but arbitrary mode is excited.
In order not to violate the energy bound,
each mode $i$ must be excited less than $l_i=\bigl[ \frac{B}{\Omega_i} \bigr]
=\bigl[\frac{E_{max}}{\hbar \omega_i} \bigr]$
times, and therefore Yurtsever derives for $S_1$ an upper bound given by
$S_1=\sum \limits_{i=1}^{N} (l_{i}-1)$.
For $S_1$, this sum can be expressed (approximately) by an integral over all allowed modes
\begin{equation}
S_1=\frac{L^3}{2 \pi^2 c^3} \int \limits_{0}^{2 \pi \mu / \tau_p} \omega^2
\frac{E_{max}}{\hbar \omega}=\frac{\sqrt{3}}{4} \mu^2 \eta \Biggl( \frac{L}{l_p} \Biggr)^4 :=z,
\end{equation}
where $l_i-1$ has been replaced by $E_{max}/\hbar \omega_i$ and the summation by an integration.
By a similar reasoning, Yurtsever derives $S_n/n! \sim z^n/(n!)^2$ for $n>1$
(eqns. (25-28) in \cite{Yurtsever}),
i.e. he looks for all solutions of eq. (\ref{start}) where the number of the non-vanishing
but arbitrary $n_i$'s is exactly $n$.

However, let us assume for the moment that $\tilde{N}$ modes are excited with
\begin{equation}
\tilde{N}=\frac{(2 \sqrt{3} \pi^2 \eta)^{3/4}}{6 \pi^2} \Biggl( \frac{L}{l_p} \Biggr)^{3/2},
\end{equation}
i.e., we focus on the term $S_{\tilde{N}}/\tilde{N}!$.
The state with the lowest possible energy is obtained by exciting all the lowest lying modes
only once. This is equivalent to filling up all modes with one quantum
up to a maximum mode frequency given by
\begin{equation}
\omega_{max}^4=\frac{2 \sqrt{3} \pi^2 \eta c^4}{L^2 l_p^2},
\end{equation}
since in correspondence with eq. (3) in \cite{Yurtsever}, a short calculation shows
that one has indeed
\begin{equation}
\tilde{N} = \frac{L^3}{2 \pi^2 c^3} \int \limits_{0}^{\omega_{max}} \omega^2 d\omega
=\frac{L^3}{2 \pi^2 c^3} \frac{\omega_{max}^3}{3},
\end{equation}
and the energy of the considered state is then given in analogy to eq. (3) in \cite{Yurtsever}
by
\begin{equation}
\tilde{E} = \frac{L^3}{2 \pi^2 c^3} \int \limits_{0}^{\omega_{max}} \omega^2 \hbar \omega d \omega
=\frac{L^3}{2 \pi^2 c^3} \frac{\omega_{max}^4}{4}=\frac{\hbar}{\tau_p} B=E_{max}.
\end{equation}
I.e., the black hole energy constraint is already reached when
$\tilde{N} \sim (L/l_p)^{\frac{3}{2}}$
different modes are excited, but the sum in eq. (\ref{toobigsum}) allows for
$N \sim (L/l_p)^3 >> \tilde{N}$ different excited modes.
Consequently, we must have $S_n/n!=0$ for $n>\tilde{N}$ in eq. (\ref{toobigsum}),
which represents a clear contradiction to the result derived by Yurtsever.
Still, eq. (28) in \cite{Yurtsever} can be used to give an upper bound for the
entropy of a constrained scalar quantum field.
Instead of 
\begin{equation}
W(B) \simeq \sum \limits_{n=0}^{N} \frac{z^n}{(n!)^2}, \, \, \mbox{with} \, \, z=\mu^2 \Biggl(
\frac{L}{l_p} \Biggr) B =\frac{\sqrt{3}}{4}\mu^2 \eta \Biggl( \frac{L}{l_p} \Biggr)^4 ,
\end{equation}
one can write
\begin{equation}
W(B) < \sum \limits_{n=0}^{\tilde{N}} \frac{z^n}{(n!)^2} < \tilde{N}
\frac{z^{\tilde{N}}}{(\tilde{N}!)^2} < z^{\tilde{N}},
\end{equation}
because $z^{\tilde{N}}/\tilde{N}!$ is the largest term appearing in the sum above
(since $z >> \tilde{N}$).
Therefore,
\begin{equation}
\log W(B) < \tilde{N} \log z < \sigma \Biggl( \frac{L}{l_p} \Biggr)^{\frac{3}{2}} \log L
\sim A^{\frac{3}{4}} \log A^{\frac{1}{4}}, \label{entrolog}
\end{equation}
where $\sigma$ is a numerical constant of order one.
A much stronger entropy bound very similar to the one given by 't Hooft $S \sim A^{\frac{3}{4}}$
is recovered. The additional, but numerically small logarithmic term is due to the fact
that the approximation eq. (27) in \cite{Yurtsever},
$S_n \sim z^n/n!$, is still too optimistic for $n$ smaller but close
to $\tilde{N}$. A more concise derivation of the entropy leads to the well-known result
given by 't Hooft $\log W(B) \sim A^{\frac{3}{4}}$, which is not presented here for
the sake of brevity. Such a derivation can be found in the meantime
in the comment \cite{Chen}, which, however, contains the erroneous statement that
in the present work it is indeed claimed that the entropy contains a logarithmic term.
This is not the case. We derived eq. (\ref{entrolog}) into order to disprove Yurtsever's
result in a most straightforward manner.

\section{The pedestrian way}
We present here another simplified explanation to highlight where
the problem with Yurtsever's combinatorics is located. $S_n/n!$
should be the number of solutions of the relation (16) in Yurtsevers's paper
[2], where exactly $n$ coefficients $n_{j_1},...,n_{j_n}$ do not vanish,
and these solutions can be interpreted as points on n-dimensional
subpolyhedra.

However, a look at Fig. 1 presented in [2] itself reveals that there is a problem already
at low dimensions. Fig. 1 shows the case $N=3$, and $l_1=B/\Omega_1=4$,
$l_2=B/\Omega_2=l_3=3$. If one count the states {\emph{inside}} the $l_2-l_3$-plane, one should obtain according to Yurtsever's formula,
$(l_3-1)(l_2-1)/2!=4/2=2$ states, but there is only one.
For the total $S_2/2!$ one obtains $(2\times2+2 \times 3 +2 \times 3)/2!=8$ according
to Yurtesever, but there are only 7 points inside the two-dimensional subpolyhedra.
Next one may consider $S_3/3!$: We have $(l_1-1)(l_2-1)(l_3-1)/3!=12/3!=2$, but there
is only {\emph{one}} point inside the polyhedron, and {\emph{not two}}.
Eq. (21) works well for $S_1$, not well for $S_2$, and fails for $S_3$.

The observed behavior is not accidental due to the low dimensional case considered,
and it is exactly displaying the problem
which becomes worse and worse at higher dimensions such that there are {\emph{no}}
allowed states anymore on subpolyhedra with dimension $n>\tilde{N}$.
Yurtsever does not take into account eq. (13) {\emph{properly}} in his paper
when calculating $S_n/n!$ as a number of points on subpolyhedra.
This seems counterintuitive, but such problems are typical for
higher-dimensional problems.

Therefore, the combinatorics of Yurtsever is tempting, but incorrect.
The observations made above are not due to 'rounding errors', if one studies
less trivial examples in higher dimensions (by counting the exact numer of states on a computer, up to the dimensions where this is possible),
one finds that indeed eq. (21) in [2] soon becomes completely useless for growing $n$.

The {\emph{irony}} of the story is that Yurtsever himself writes after eq. (16) in
[2]: "One might be tempted to conclude that W(B) is simply proportional
to the volume $\mathcal{P}^N$...". But then, he uses a completely analogous
{\emph{volume}} formula eq. (21) to calculate the number of points in high-dimensional
subpolyhedra.
{\emph{Not only}} the $\mathcal{P}^N$ is problematic, but also most of the
higher-dimensional $\mathcal{P}^n$'s. Yurtsever's formula is only useful at low
dimensions (which are {\emph{irrelevant}} concerning their contribution
to the full $W(B)$ according to Yurtsever, but which generate the much smaller
true entropy \`a la 't Hooft).

In this paper, an example was presented where {\emph{only}}
lowest lying states of the scalar field are occupied in order to show that
it is indeed {\emph{impossible to find points}} on subpolyhedra with dimension
$> \tilde{N}$, without violating the Black hole energy bound.

\section{Conclusions}
The derivation of a holographic entropy bound by counting states of a system of
free scalar bosons enclosed within a spatially bounded region in flat spacetime
presented in \cite{Yurtsever} and generalized \emph{bona fide} in \cite{Aste}
to arbitrary spacetime dimensions is flawed by a mathematical error. It should also be
pointed out that the correct $A^{\frac{3}{4}}$ result for the entropy has been derived
recently within a different framework in \cite{Hsu}.

Although a holographic entropy bound for a bosonic system in flat spacetime
would have been an interesting result, one should keep in mind that in order to
establish the existence of a holographic bound stemming from gravitational stability,
one should work within a general relativistic quantum field theory setting.
It is well possible that the computation of the dimensionality of a scalar field Hilbert
space will then lead to a different result than in flat space, and any framework not
employing general relativistic (differential geometric) methods is likely to produce
unreliable results.

\section{Acknowledgment}
The author wishes to thank Stephen Hsu for valuable discussions.
This work was supported by the Swiss National Science Foundation.

\end{document}